\documentclass[12pt,a4paper,final]{iopart}
\usepackage{iopams}  
\usepackage{graphicx}
\usepackage{epsfig}
\usepackage{epstopdf}
\usepackage[breaklinks=true,colorlinks=true,linkcolor=blue,urlcolor=blue,citecolor=blue]{hyperref}

\begin{document}

\title[Population trapping in the excited states]{Population trapping in the excited states using vacuum-induced coherence and adiabatic process}

\author{Babu Lal Kumawat$^1$, Pardeep Kumar$^2$, and Shubhrangshu Dasgupta$^1$ }
\address{$^1$Department of Physics, Indian Institute of Technology Ropar, Rupnagar, Punjab 140001, India}
\address{$^2$School of Physics and Astronomy, Rochester Institute of Technology, 84 Lomb Memorial Drive, Rochester, NY 14623, USA}
\eads{\mailto{babu.lalkumawat@iitrpr.ac.in}, \mailto{sdasgupta@iitrpr.ac.in}}

\begin{abstract}
We theoretically investigate how population can be trapped  in the excited doublet in presence of vacuum-induced coherence (VIC). We employ delayed pulses to transfer population from a metastable state to the excited states. Subsequently, spontaneous emission from the excited doublet builds coherence between them. This coherence can be probed by using chirping,  which leads to the decoupling of the excited doublet from the ground state thereby ensuring population transfer via delayed pulses. Our results indicate that the existence of VIC leads to the generation of a mixed state in the excited state manifold, where trapping of the population occurs even in the presence of large decay. This trapping may be realized in molecular systems and can be understood as a sensitive probe of VIC. We present suitable numerical analysis to support our results. 

\end{abstract}

\vspace{2pc}\noindent{\it Keywords}: vacuum-induced coherence, adiabatic transfer, delayed pulses, chirping, purity
\maketitle
\newpage
\section{Introduction}

Quantum coherence \cite{Ficek2007} in atomic/molecular systems is the backbone of coherent control of optical properties of the medium. Such coherence arises from the interaction between a coherent source of light and the medium.  Beside this, incoherent process such as spontaneous emission also leads to coherence between the excited doublets in the medium.  One of the pronounced coherence phenomena called, \textit{vacuum-induced coherence} (VIC), arises due to  interference between different pathways of spontaneous emission from the excited state doublet to a single ground  state \cite{Agarwal1974}. In recent years, many atomic systems are studied for VIC to show fascinating applications \cite{Harris1989,Scully1989,Zhou1997,Keitel1999,Sunish1998,Emmanuel1998,Knight1998}. Furthermore, several models have been proposed to recognize  VIC in ions \cite{Das2008,Kiffner2006}, quantum dots \cite{Gurdev2005,Economou2005} and M\"{o}ssbauer nuclei \cite{Heeg2013}. However, the occurrence of VIC requires a stringent condition that the dipole moments of participating transition should be non-orthogonal \cite{Agarwal1974,imamoglu1989}. The unavailability of this condition in atomic systems restricts their applications. Contrary to this, molecular systems are considered to be favorable candidates for the realization of VIC. In fact, VIC arises naturally in molecules due to availability of excited doublet belonging to same molecular electronic state thereby satisfying the requirement of non-orthogonality. Recently, the existence of VIC is reported in the excited ro-vibrational states of an atom-molecule system by means of ultracold photoassociative techniques \cite{Deb2012}.  Furthermore, it is observed that  in ultracold molecules  VIC can be detected and estimated  using magneto-optical rotation \cite{Kumar2016}.

Population trapping in excited states requires suitable initial prepration of the system. One possible way of transferring prepartion in trapped state could be to transfer population from ground state by using a `$\pi$' pulse. However, this would also create a coherence between the excited states and ground state. To avoid, creation of such coherence, we employ the method stimulated Raman adibatic process (STIRAP). Originally STIRAP was introduced as an efficient way to transfer the population between the dipole forbidden levels in multilevel systems \cite{Gaubatz1990}.  Using this technique, Shore and his co-workers showed that one can transfer population to a deserved energy level, without populating the intermadiate levels \cite{Shore1991}. The STIRAP process have been extensively studied for numerous applications ranging from quantum optics \cite{Esslinger1996, Kulin1997, Parkins1993} to quantum chemistry \cite{Dittmann1992} .  

In this paper, we show how to adiabatically transfer the population to an excited doublet, exhibiting VIC, from a third excited state, thanks to STIRAP. The third excited state can be chosen as a meta-stable state. Contrary to usual STIRAP, which involves two or more ground states, we rather employ this technique to transfer population  between  a manifold of excited states. Once initially prepared, the subsequent spontaneous emission from the excited doublet to common ground state creates VIC. It should be borne in mind that such VIC is shown to lead to a coherent superposition of the excited states in a V-type configuration during \textit{photoassociation} of cold molecules \cite{Deb2012}. Moreover, under suitable conditions, the superposition state of the excited doublet can become a dark state (immune to the relaxation) where initially transferred population gets trapped \cite{Deb2012}.  In this paper, we propose an alternative approach using STIRAP to create such a superposition state (albeit partially pure) at steady state, in presence  of  dissipative channels. We specifically show that it is possible to trap population in the excited state doublet coherently, with the aid of VIC and using the technique STIRAP. We analyze our result by  investigating the purity of the state thus prepared.

The paper is organized  as follows: In section 2 we describe the theoretical  model with the dynamical evolution including VIC. In Section 3 we present the numerical analysis by considering different cases. We  conclude the paper in section 4.

\section{Model}

We consider a molecular system whose spin-singlet ($S=0$) ground state can be represented as $\vert 0\rangle=\vert v, J=0, m_J, S=0, L=0, M_L=0\rangle$, where $v$ is the vibrational level,  $L$ ($J$) is electronic orbital angular momentum  (rotational quantum number) and $M_L$ ($m_{J}$) is the projection of angular momentum (rotational level) along the internuclear axis which is molecule-fixed z-axis. In same manner, we choose near degenerate triplet excited states $\vert 1\rangle=\vert v^{\prime}, J=1, m_J=1, S=0, L=1, M_L=+1\rangle$ and $\vert 2\rangle=\vert v^{\prime}, J=1, m_J=1, S=0, L=1, M_L=-1\rangle$, where $v^{\prime}$ represents the vibrational quantum numbers of the excited states. Both of these states belong to  $^{1}\Pi$ state. Note that the degeneracy in the excited states will be lifted due to the orbit-rotation coupling \cite{Landau1997}. However, this lifting would be very small as compared to spontaneous linewidth \cite{Agarwal1974}. Further, we should mention that the two transitions considered here occur between the same rotational and electronic states. Therefore, the electronic part of the dipole matrix elements will be the same thereby making the transition dipole moments $\vec{d}_{01}$ and $\vec{d}_{02}$ parallel \cite{imamoglu1989}. Thus, the proposed molecular levels ultimately satisfy the conditions for VIC to occur.  A meta-stable state $\vert 3\rangle=\vert v_1, J=1, m_J=-1, S=0, L=1, M_L=0\rangle$ is taken such as the transition dipole $\vec{d}_{03}$ will be perpendicular to $\vec{d}_{01}$ and $\vec{d}_{02}$. 

\begin{figure}[htb!]
\centering
\includegraphics[scale=0.18]{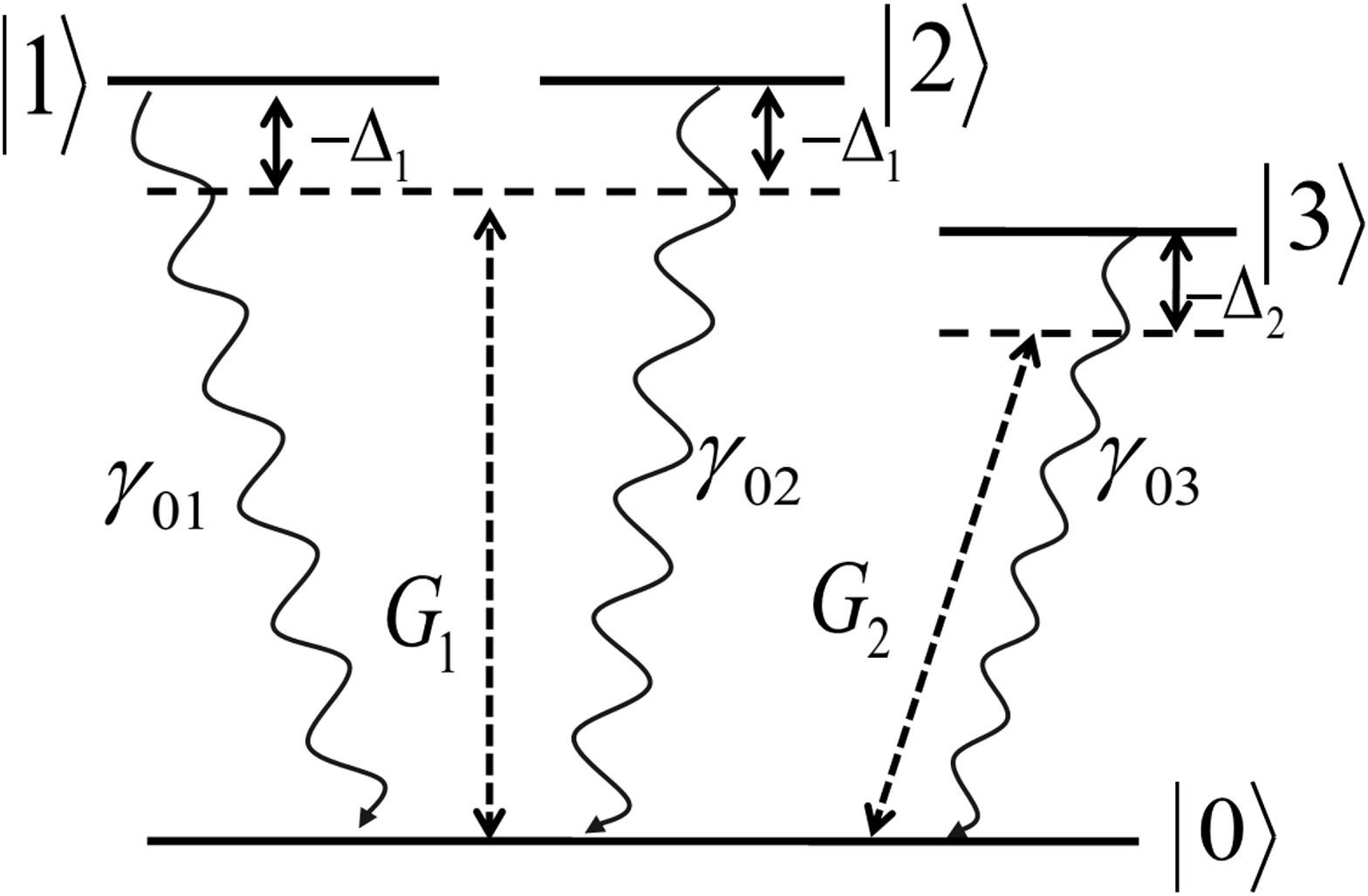}
\caption{Systematic energy level diagram of four level system. The levels $|1\rangle$ and $|2\rangle$ are coupled to the ground state $|0\rangle$ by $\hat{x}$-polarized field with Rabi frequency $G_{1}$. A $\hat{y}$-polarized field of Rabi frequency $G_{2}$ drives the transition $|0\rangle\leftrightarrow|3\rangle$. Here $\gamma_{0i}$ ($i$=1,2,3) is the spontaneous decay rate of level $\vert i\rangle$ and, $\Delta_1$ and $\Delta_2$ are the detuning of corresponding coupled field.}
\label{fig1}
\end{figure}

Now, we consider a four level configuration as shown in figure \ref{fig1},  in which the ground state  $ \vert 0 \rangle $ is coupled to the excited doublet $\vert 1 \rangle$ and $\vert 2 \rangle$ by an electromagnetic field $\vec{E_1}$. Another excited state  $\vert 3 \rangle$ is coupled to the ground state $\vert 0\rangle$ via an electromagnetic field $\vec{E_2}$. These electromagnetic fields are represented as

\begin{eqnarray}
& & \vec{E}_1(t)=\hat{x} \varepsilon_1(t) e^{-i\omega_{G_1}t-i\phi_1 (t)}+c.c., \nonumber \\
 & & \vec{E}_2(t)=\hat{y} \varepsilon_2(t)e^{-i\omega_{G_2}t-i\phi_2 (t)}+c.c..
\end{eqnarray}
where, $\varepsilon_{1,2}(t)$ are the time-dependent amplitudes of the electromagnetic fields and the polarization of the pulses  are perpendicular to each other. The time-dependent phases  $\phi_{1,2} (t)$ denote frequency-sweep or chirping.

The Hamiltonian of this system under the dipole approximation can be written as

\begin{eqnarray}
\hat{H} &= &\hbar[\omega_{10}\vert 1\rangle \langle 1\vert +\omega_{20}\vert 2\rangle \langle 2\vert +\omega_{30}\vert 3\rangle \langle 3\vert] \nonumber \\
&  &-[(\vec{d}_{10} \vert 1\rangle \langle 0\vert + \vec{d}_{20} \vert 2\rangle \langle 0\vert +H.c.).\vec{E}_1] -[(\vec{d}_{30} \vert 3\rangle \langle 0\vert + H.c.).\vec{E}_2]\;,
\end{eqnarray}
Here, zero of the energy is defined at the level $|0\rangle$ and $\hbar\omega_{\alpha\beta}$ is the energy difference between $|\alpha\rangle$ and $|\beta\rangle$. Now, the dynamical evolution of the system can be described by the following density matrix equations under the rotating wave approximation:

\begin{eqnarray}
\dot{\tilde{\rho}}_{11} &=& -\gamma_{01}\tilde{\rho}_{11}-\frac{\gamma_{12}}{2}(\tilde{\rho}_{12}+\tilde{\rho}_{21})+i[G_1(t)\tilde{\rho}_{01}+G_{1}^{*}(t)\tilde{\rho}_{10}] , \nonumber \\
\dot{\tilde{\rho}}_{22}&=&-\gamma_{02}\tilde{\rho}_{22}-\frac{\gamma_{12}}{2}(\tilde{\rho}_{12}+\tilde{\rho}_{21})+i[G_1(t)\tilde{\rho}_{02}+G_{1}^{*}(t)\tilde{\rho}_{20}] , \nonumber \\
 \dot{\tilde{\rho}}_{33}&=&-\gamma_{03}\tilde{\rho}_{33}+i[G_2(t)\tilde{\rho}_{03}+G_{2}^{*}(t)\tilde{\rho}_{30}] , \nonumber \\
\dot{\tilde{\rho}}_{21}&=&-i \Gamma_{21}\tilde{\rho}_{21}-\frac{\gamma_{12}}{2}(\tilde{\rho}_{11}+\tilde{\rho}_{22})+i[G_1(t)\tilde{\rho}_{01}-G_{1}^{*}(t)\tilde{\rho}_{20}] , \nonumber \\
\dot{\tilde{\rho}}_{10}&=& \left[i\Delta_1(t)-\Gamma_{10}\right]\tilde{\rho}_{10}-\frac{\gamma_{12}}{2}\tilde{\rho}_{20}+iG_1(t)(1-2\tilde{\rho}_{11}-\tilde{\rho}_{22}-\tilde{\rho}_{33}) \nonumber \\
&&-iG_1(t)\tilde{\rho}_{12}-iG_2(t)\tilde{\rho}_{13} , \nonumber \\
\dot{\tilde{\rho}}_{20}&=&\left[i\Delta_1(t)-\Gamma_{20}\right]\tilde{\rho}_{20}-\frac{\gamma_{12}}{2}\tilde{\rho}_{10}+iG_1(t)(1-\tilde{\rho}_{11}-2\tilde{\rho}_{22}-\tilde{\rho}_{33}) \nonumber \\
&&-iG_1(t)\tilde{\rho}_{21}-iG_2(t)\tilde{\rho}_{23} , \nonumber \\
\dot{\tilde{\rho}}_{30} &=& \left[i\Delta_2(t)-\Gamma_{30}\right]\tilde{\rho}_{30}+iG_2(t)(1-\tilde{\rho}_{11}-\tilde{\rho}_{22}-2\tilde{\rho}_{33}) \nonumber \\
&&-iG_1(t)\tilde{\rho}_{31}-iG_1(t)\tilde{\rho}_{32} , \nonumber \\
\dot{\tilde{\rho}}_{31}&=&\Big[i\Big(\Delta_2(t)-\Delta_1(t)\Big)-\Gamma_{31}\Big]\tilde{\rho}_{31}-\frac{\gamma_{12}}{2}\tilde{\rho}_{32}+iG_2(t)\tilde{\rho}_{01}-iG_{1}^{*}(t)\tilde{\rho}_{30} , \nonumber \\
\dot{\tilde{\rho}}_{32}&=&\Big[i\Big(\Delta_2(t)-\Delta_1(t)\Big)-\Gamma_{32}\Big]\tilde{\rho}_{32}-\frac{\gamma_{12}}{2}\tilde{\rho}_{31}+iG_2(t)\tilde{\rho}_{02}-iG_{1}^{*}(t)\tilde{\rho}_{30}\;,
  \end{eqnarray}
where, $\gamma_{12}=\sqrt{\gamma_{01}\gamma_{02}}\cos \theta$, is the cross coupling term between transition $\vert 1\rangle \leftrightarrow \vert 0\rangle$ and $\vert 2\rangle \leftrightarrow \vert 0\rangle$ and $\cos \theta =\frac{\vec{d}_{10}.\vec{d}_{20}}{\vert \vec{d}_{10}\vert \vert \vec{d}_{20}\vert}$ with $\theta$ being the angle between the dipole moment elements $\vec{d}_{10}$ and $\vec{d}_{20}$. The existence of this term reflects that the same vacuum mode is interacting with both the transitions, thereby giving rise to VIC between the two levels $\vert 1\rangle $ and $\vert 2\rangle$. Here, $G_i(t),(i=1,2)$ represents time-dependent Rabi frequency of corresponding electromagnetic field.  $\Gamma_{ij}=\frac{1}{2}\sum\limits_k(\gamma_{ki}+\gamma_{kj})+\gamma_{coll}$ is the decay rate of the coherence  between level $\vert j \rangle$ and $\vert i\rangle$, and $\gamma_{coll}$ is the collisional decay rate.  Further, $\Delta_1(t)=\omega_{G1}+\dot{\phi}_1(t)-\omega_{10} $,  and $\Delta_2(t)=\omega_{G2}+\dot{\phi}_2(t)-\omega_{30}$ are the detuning between the frequency of the  field and the  atomic transition frequency, where $\dot{\phi}_i(t)=\chi_i F_i(t), (i=1,2)$ represents chirping (sweeping) in frequency with chirping constant $\chi_i$ and time-dependence $F_i(t)$ describes the chirping profile \cite{Asoka}. The transformation used to drive the density matrix elements in (3) are as follows: $\rho_{01}=\tilde{\rho}_{01}e^{-i\omega_{G_{1}}t}$, $\rho_{02}=\tilde{\rho}_{02}e^{-i\omega_{G_{1}}t}$, $\rho_{03}=\tilde{\rho}_{03}e^{-i\omega_{G_{2}}t}$, and rest of the elements remain same. Also, the above density matrix elements satisfy the conditions $\tilde{\rho}_{00}+\tilde{\rho}_{11}+\tilde{\rho}_{22}+\tilde{\rho}_{33}=1$ and $\tilde{\rho}_{ij}=\tilde{\rho}_{ji}^{*}$.

  \section{Numerical analysis}
  As discussed in the introduction, we choose to prepare the molecule in a third excited state $\vert 3\rangle$, which can be chosen to be a meta-stable state. One can apply pulses in suitable sequence such that the population is transferred (albeit partially) to the excited doublet. We consider  Gaussian profile for the time-dependence of the pulses
   
   \begin{eqnarray}
   G_1(t)&=G_{01}\exp (-t^2/\tau^2)\;,  \nonumber \\
G_2(t)&=G_{02}\exp (-(t-t_0)^2/\tau^2)\;,
   \end{eqnarray}
   where, $t_0$ is the pulse delay,  $\tau$ in the pulse width and $G_{01}$ and $G_{02}$ are the maximum amplitude of corresponding pulse. Here we consider $G_1(t)$  as a Stoke pulse and $G_2(t)$  as a pump pulse so that pump pulse precedes the Stokes pulse. This is akin to the pulse sequence, as employed in STIRAP for population transfer between ground states.

\subsection{Case I: Without chirping}
\begin{figure}[htb!]
\centering
 \includegraphics[scale=.4]{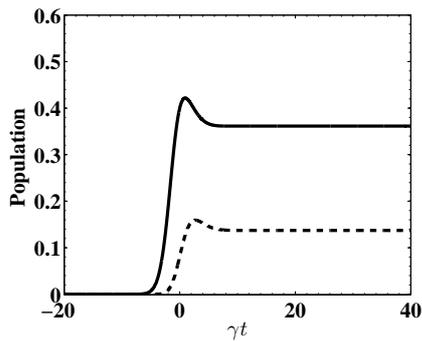}
\caption{Population variation of level $\vert 1\rangle$ and $\vert 2\rangle$ with respect to time, where dashed(solid) line representing population of excited state $\vert 1\rangle$ ($\vert 2\rangle$), angle between transition dipole  $\vec{d}_{10}$ and $\vec{d}_{20}$ is $\theta=0$, $G_{01}=0.9\gamma$, $G_{02}=0.3\gamma$, pulse width $\tau=4/\gamma$, pulse delay time $t_0=2.5\tau$,  and spontaneous decay rates $\gamma_{01}=5.8\gamma$, $\gamma_{02}=2.2\gamma$ and $\gamma_{03}=0.1\gamma$. }
\label{fig2}
\end{figure}

Next, we solve  (3) using (4). We first consider that the interaction of the fields with the system at exact resonance $\left(\Delta_i=0\right)$  without chirping. Also, we consider a situation where maximum VIC exists in the system (i.e. $\theta=0$). We display the evolution of the population of the level $\vert 1\rangle$ and $\vert 2\rangle$  in figure \ref{fig2}.
Clearly, almost 50\% population is transferred to the excited doublet and  gets  trapped at steady state. The transferred population gets confined  in excited doublet manifold that refers to a coherent (albeit partially) superposition state of excited state arising due to VIC.  

To investigate further, we show the variation of real and imaginary part of coherence  $\tilde{\rho}_{10}$, $\tilde{\rho}_{20}$, $\tilde{\rho}_{31}$ and $\tilde{\rho}_{32}$ with respect to time.

\begin{figure}[htb!]
\centering
 \includegraphics[scale=.45]{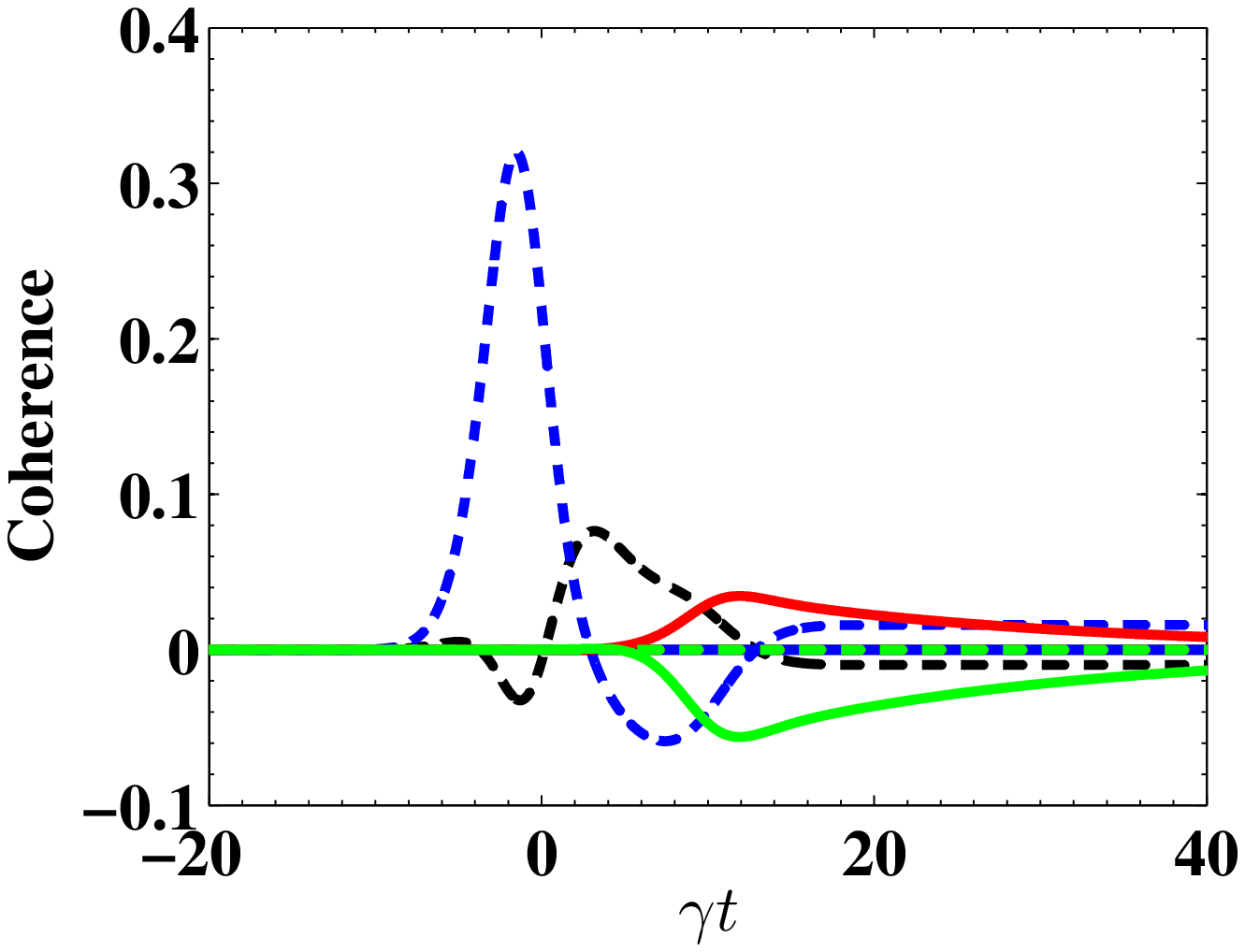}
\caption{(Color online) Variation of real and imaginary part of  coherence $\tilde{\rho}_{10}$(black line), $\tilde{\rho}_{20}$(blue line), $\tilde{\rho}_{31}$(red line) and $\tilde{\rho}_{32}$(green line) with respect to time at exact resonance($\Delta_i=0$), here solid(dashed) lines represents the real(imaginary) part of  coherence  and  other common parameters are same as in figure \ref{fig2}.}
\label{fig3}
\end{figure}

 We found that the magnitude of these coherence are non-zero that means that the density matrix of the system does not get factorized into the basis $\{\vert 1\rangle$, $\vert 2\rangle \}$. This further refers to a coherent signature of the ground state in the final density matrix. We ask the following question next: Is it possible to trap the population maximally exclusively in the excited doublet, assisted by VIC ? In the following, we find an affirmative answer.
 
\subsection{Case II: With chirping }
To eliminate coherence between the excited doublet and the ground state, we next employ chirped pulses as follow:
\begin{eqnarray}
\Delta_1(t)&=\chi_1\tanh(t)\;, \nonumber \\
\Delta_2(t)&=\chi_2\tanh(t-t_{0})\;.
\end{eqnarray}

\begin{figure}[htb!]
\centering
 \includegraphics[scale=.45]{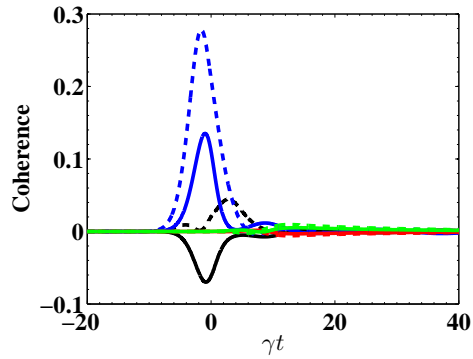}
\caption{(Color online) Variation of real and imaginary part of  coherence $\tilde{\rho}_{10}$(black line), $\tilde{\rho}_{20}$(blue line), $\tilde{\rho}_{31}$(red line) and $\tilde{\rho}_{32}$(green line) with respect to time for $\Delta_i \neq 0$ at $\chi_1=0.3$ and $\chi_{2}=0.2$, here solid(dashed) lines represent the real(imaginary) part of  coherence. Other common parameters are same as figure \ref{fig3}.}
\label{fig4}
\end{figure}

We display the real and imaginary part of coherence  $\tilde{\rho}_{10}$, $\tilde{\rho}_{20}$, $\tilde{\rho}_{31}$ and $\tilde{\rho}_{32}$  in figure \ref{fig4} for $\Delta_i \neq 0$.

We find that for suitable choice of $\chi_i$, all real and imaginary parts of these coherences  become zero. Therefore, the excited states are no longer coupled with the ground state and the meta-stable state. This  means  that population gets transferred from the  meta-stable to the excited states via delayed pulses.

So far, we discussed about how the population gets  transferred to the excited doublet. Now, we will discuss, why this population gets confined at excited states for longer time?  To find the answer of this question we calculate coherence $\tilde{\rho}_{21}$  as discussed next
\begin{figure}[htb!]
\centering
 \includegraphics[scale=.45]{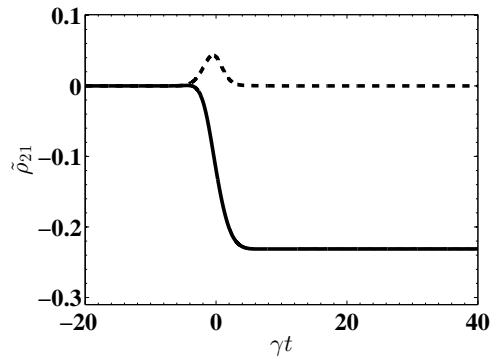}
\caption{Variation of real and imaginary part of  coherence $\tilde{\rho}_{21}$ with respect to time, here solid(dashed) line represents the real(imaginary) part of  coherence  and  other common parameters are same as figure \ref{fig4}.}
\label{fig5}
\end{figure}
 
\begin{figure}[htb!]
\centering
 \includegraphics[scale=.45]{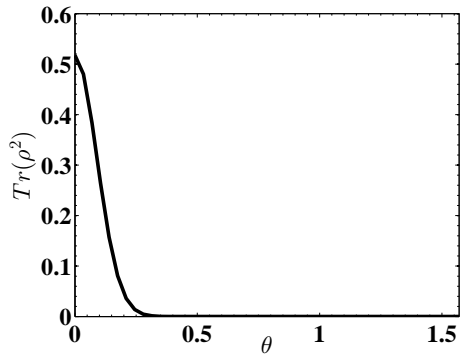}
\caption{Variation of purity with respect to $\theta$ at steady state, where $\theta$ is  the angle between transition dipole  $\vec{d}_{10}$ and $\vec{d}_{20}$. Rest of the parameters are same as in figure \ref{fig5}.}
\label{fig6}
\end{figure}

 The time-dependent real and imaginary part of  coherence $\tilde{\rho}_{21}$  are shown in figure \ref{fig5} Clearly, the magnitude of this coherence has non-zero value. Since, there is no transition amplitude between the excited states and the transition dipole moments $\vec{d}_{10}$ and $\vec{d}_{20}$ are parallel to each other. The cross coupling term $\gamma_{12}$ has non-zero value  that means spontaneous emission channels of excited states make a cross-talk via the vacuum of electromagnetic field and this interaction is responsible for VIC between excited states. It is interesting to see that the VIC causes to trap the population in the excited doublet, even in the presence of large spontaneous emission rate. This trapping of population in the excited doublet may be considered as a probe of VIC. Further, to understand the role of VIC, we continue to investigate the purity $Tr(\rho^2)$ of the state thus prepared. Here $\rho$ designates the density matrix block in the $\{\vert 1\rangle$, $\vert 2\rangle$ subspace only at the steady state. As $\theta$ increases to $\pi /2$, the VIC ceases to exit. We show in figure \ref{fig6} how the purity changes with $\theta$.

  It is clear that in the presence of VIC (corresponding to $\theta=0$), the purity becomes maximum and zero for $\theta \gtrsim 0.3$. Note that in the present case, the system is prepared in a mixed state at long times. 
  
\section{Conclusion}
In conclusion, we have described the transfer of population from meta-stable state to excited doublet via delayed pulses. Further, chirped detuning makes sure that the population is transferring only due to pulse. Once initially prepared, a coherence in the excited doublet builds up due to the  interaction of the transition dipole moments with the vacuum of electromagnetic field. We have shown that due to this coherence the transferred population is trapped at excited states at steady state. We verified  the effect of VIC and the pulse through numerical analysis of time evolution and purity.

\end{document}